# Non-local low energy neutral excitations in a strongly disordered triangular Mott magnet $Cr_3Se_2Br_5$

Wenhao Liu[1,2,3,†], Dechen Zhang[4], Yuanqi Lyu[1], Lebing Chen[1,2], Lifang Hu[5], Keith M. Teddei[6], Yuting Zhang[4], Steve Shelton[10], Moon Kim[5], Xiqu Wang[7], Michael A. Susner[11], James G. Analytis[1,2,8,9], Dung-Hai Lee[1,2], Lu Li[4], Bing Lv[3,†] & Robert J. Birgeneau[1,2,†]

1. Department of Physics, University of California, Berkeley, CA, USA
2. Material Sciences Division, Lawrence Berkeley National Lab, Berkeley, CA, USA
3. Department of Physics, the University of Texas at Dallas, Richardson, TX, USA
4. Department of Physics, University of Michigan, Ann Arbor, MI, USA
5. Department of Materials, the University of Texas at Dallas, Richardson, TX, USA
6. Neutron Scattering Division, Oak Ridge National Laboratory, Oak Ridge, TN, USA
7. Department of Chemistry, University of Houston, Houston, TX, USA
8. CIFAR Quantum Materials, Toronto, ON, Canada
9. Kavli Energy NanoScience Institute, Berkeley, CA, USA
10. Molecular Foundry, Lawrence Berkeley National Laboratory, Berkeley, CA, USA
11. Materials and Manufacturing Directorate, Air Force Research Laboratory, Wright Patterson Air Force Base, OH, USA.

**Understanding if low-energy excitations can remain itinerant in the presence of strong disorder remains a central challenge in frustrated quantum magnets, where disorder is generally expected to localize excitations through Anderson-like mechanisms. Here we report the emergence of charge-neutral itinerant excitations in a van der Waals compound $Cr_3Se_2Br_5$, a strongly disordered $S$ = 3/2 Mott insulator with a frustrated triangular lattice. Structural analysis reveals substantial intrinsic disorder arising from Cr-site deficiency and Se/Br-site mixing, which appears to be fixed and cannot be readily tuned. No long-range magnetic order or conventional glassy behavior is observed. In addition to its highly insulating nature, the magnetic specific heat $C_{mag}/T$ and thermal conductivity $\kappa_{xx}/T$ both exhibit linear temperature dependencies with substantial finite intercepts. In particular, a sizeable field-independent residual term $\kappa_0/T \approx 0.03$ W m$^{-1}$ K$^{-2}$ is observed as $T \to 0$, providing compelling evidence of itinerant low-energy excitations that carry entropy without charge. These findings conceptually advance our understanding of quantum matter by demonstrating a rare regime where the interplay of disorder, frustration, and electronic correlations actively reshapes the nature of low-energy excitations, allowing itinerant neutral excitations to coexist with strong intrinsic disorder.**

†Corresponding authors, E-mail: wenhao.liu@berkeley.edu; blv@utdallas.edu; robertjb@berkeley.edu

Collective quantum phenomena lie at the heart of modern condensed matter physics, often manifesting as emergent quasiparticles that enrich our understanding of many-body systems[1,2] . In conventional metals, low energy excitations are described by itinerant electronic quasiparticles within Landau's Fermi liquid framework, allowing charge and heat transport to persist down to the zero-temperature limit. In contrast, in insulators, especially for Mott insulators, charge degrees of freedom are frozen by strong Coulomb interactions. As a result, low-energy excitations are conventionally dominated by charge-neutral bosonic modes such as phonons or magnons, whose contributions to thermal transport typically diminish toward the zero-temperature limit. A long-standing question is whether insulating systems can host itinerant, gapless low-energy excitations despite the absence of charge transport when $T \rightarrow 0$ K. If present, such excitations must be electrically neutral, representing a fundamentally distinct class of emergent quasiparticles beyond conventional charged carriers.

On the theoretical side, several frameworks have been proposed in this regime, spanning from collective bosonic modes to fractionalized fermionic quasiparticles and gauge-coupled excitations[3,4]. Among these proposals, one of the most influential classes is the resonating valence bond quantum spin liquid (QSL) state[5–8] , which has gained much attention due to its similarity to high-temperature superconductivity[9,10]. In an ideal QSL state, spins remain strongly correlated and continue to fluctuate down to the lowest temperatures without symmetry breaking or the formation of any static long-range magnetic order[6,7,11–13]. The elementary excitations are fractionalized quasiparticles known as spinons, which carry spin without electric charge and may remain itinerant down to zero-temperature limit[14]

Despite many theoretical and experimental efforts over several decades, direct experimental evidence for itinerant charge-neutral excitations remains remarkably scarce. Detecting such charge neutral low-energy excitations is challenging because they couple weakly to standard experimental charge transport and electrodynamics probes, often lead to ambiguous experimental interpretations. In detecting spinons in QSLs, inelastic neutron scattering has played a central role by directly probing spin dynamics to reveal broad excitation continua expected from fractionalized quasiparticles. However, the interpretation of these signatures is often complicated, since similar continuum responses can also arise from disorder, short-range correlations, or exchange randomness[15,16] . Another technique, low-temperature thermal transport measurement, can provide a highly sensitive and complementary probe of itinerant charge-neutral quasiparticles, since a finite residual linear term in the thermal conductivity $\kappa_0/T$ at low temperatures is widely regarded as evidence for itinerant gapless excitations[13] . However, in many proposed QSL candidates, the residual $\kappa_0/T$ remains extremely small or even absent, and in some systems conflicting results have been reported [17,18]. These observations highlight the complexity in this field and continuously fuel ongoing debates regarding the nature of ground states.

Beyond complicating the experimental interpretation, disorder itself may provide an alternative route toward unconventional quantum states. In correlated materials, disorder has been closely intertwined with a variety of emergent quantum phenomena where anomalous transport emerges, including metal–insulator transitions, unconventional superconductivity, and non-Fermi-liquid behavior.[19–21] In the context of QSL, disorder is often expected to suppress quantum coherence, promote glassy or ordered states and localize the low-energy excitations via Anderson-like mechanisms.[16,22] More recent theoretical studies have suggested that disorder may also stabilize exotic phases such as random-singlet phases[23,24] or disorder-induced quantum spin liquids[25,26]. In these scenarios, disorder introduces a subtle competing tendency: it can generate itinerant low-energy excitations while simultaneously localizing them[27–29]. As a result, whether such low-energy excitations can remain itinerant in the zero-temperature limit remains an important unresolved question, and direct experimental evidence for mobile neutral excitations in strongly disordered frustrated magnets remains remarkably scarce. These considerations raise the intriguing possibility that intrinsic disorder may play a far more active role in correlated insulators than previously anticipated, potentially enabling resilient, unconventional itinerant neutral excitations to emerge

In light of these open questions, we report a new strongly disordered triangular-lattice Mott insulator $Cr_3Se_2Br_5$ (*S*=3/2), which is built from trigonal-type transition metal dichalcogenide (TMD) layers. Structure refinement and chemical analysis indicate ~15% Cr-site deficiency together with homogeneous Se/Br mixing at all crystallographic Wyckoff sites, yielding a homogeneously disordered triangular-lattice network. Unlike conventional disorder, these structural features are inherent to the material and cannot be continuously tuned as they are necessary to charge balance in the overall structure. $Cr_3Se_2Br_5$ is highly electrical insulating with a resistivity greater than $10^8$ $\Omega \cdot$ cm. Despite its odd electron count and partially filled electronic structure, optical absorption measurements reveal a sizable charge gap of approximately 1.35 eV, indicating a correlation-driven Mott insulating state. Magnetic susceptibility shows no evidence of long-range magnetic order at low temperatures, while AC susceptibility and thermoremanent magnetization (TRM) measurements further exclude typical spin-glass behavior. Strikingly, despite the large spin state ($S$ = 3/2) which typically favors semi-classical ordering, the low-temperature magnetic specific heat exhibits substantial low-energy spectral weight while the thermal conductivity $\kappa_{xx}/T$ exhibitlinear temperature dependencies as $T \to 0$. In particular, thermal conductivity measurements reveal a sizeable residual linear term $\kappa_0/T \approx 0.03$ W m$^{-1}$ K$^{-2}$, consistent with itinerant neutral excitations that remain mobile toward the zero-temperature limit. Taken together, these results suggest $Cr_3Se_2Br_5$ as a platform in which itinerant neutral excitations emerge in the presence of strong intrinsic disorder on a frustrated lattice, pointing to an unusual regime of quantum materials where the interplay of disorder, frustration, and electronic correlations fundamentally reshapes low-energy excitations

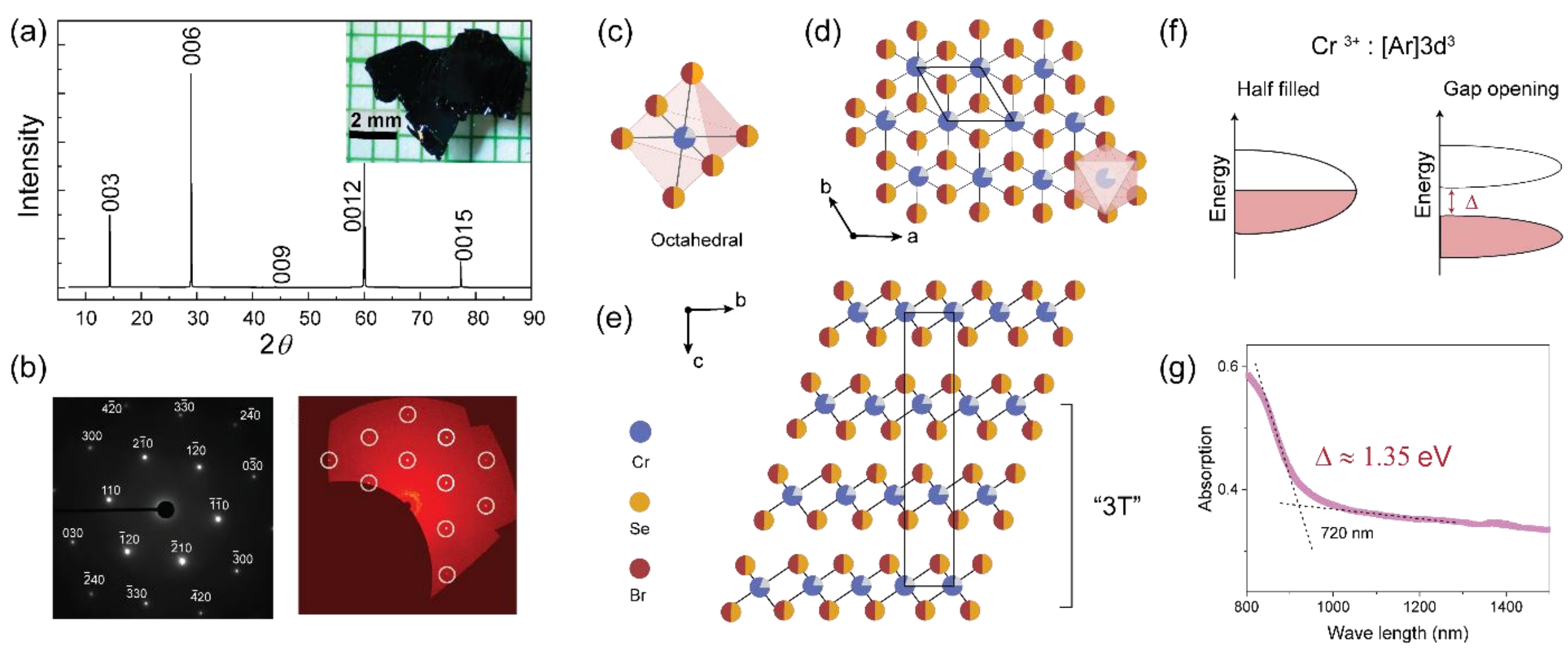

Fig. 1 **Structural and electronic characterization of $Cr_3Se_2Br_5$.** (a) X-ray diffraction on single crystals. Inset: optical image of a single crystal. (b) Electron diffraction patterns with zone axis of [001] and precession image of $Cr_3Se_2Br_5$ obtained from X-ray diffractions in the zone of (*hk0*). (c)(d)(e) The crystal structure of $Cr_3Se_2Br_5$, highlighting 3T layered features with 15% Cr deficiency and homogeneous mixture on Se and Br on the corner sites. (f) Illustration of the gap opening in a half-filled band (g) Optical absorption spectrum of $Cr_3Se_2Br_5$.

Fig.1a shows the X-ray diffraction (XRD) pattern of a $Cr_3Se_2Br_5$ single crystal with Miller indices. Only (*00l*) peaks can be observed, as expected, indicating that the flat surface is perpendicular to the crystallographic *c*-axis. The inset of Fig. 1c shows an optical image of a $Cr_3Se_2Br_5$ single crystal on a millimeter grid. The black and shiny single crystals with typical lateral sizes of 5 mm ×7 mm can be obtained. Fig.1b shows an electron diffraction pattern along the [001] zone axis together with X-ray reconstructed reciprocal-space precision images for the (*hk0*) plane. Sharp Bragg diffractions are observed without distortions, ring-like residues, or blurry spots, confirming the high single crystallinity of the sample. From the selected area diffraction pattern, the first- and second-order reflections correspond to d-spacings of $d_1$ = 1.81 Å and $d_2$ = 1.04 Å. The interplanar angle between the first-order reflection planes is 60°, while the angle between the first- and second-order reflection planes is 30°, consistent with the expected symmetry of the lattice. Combining electron diffraction with single-crystal XRD, the structure is refined in the trigonal space group R-3. Detailed crystallographic refinement details and atomic coordinates are summarized in Supplementary Table S1 and S2.

The crystal structure is shown in Fig. 1c-e. The unit cell consists of three trigonal layers, with each adjacent layer shifted by 1/3 of the in-plane lattice vector. Thus, we denote this structure as "3T". Cr and Se/Br form octahedra building units, with Cr at the centers and Se/Br at the corners. The octahedra are edge-shared forming trigonal layers with a geometrically frustrated Cr triangular lattice. The layers are stacked via weak van der Waals interactions, enabling the easy exfoliation of thin flakes from the bulk crystals shown in Supplementary Fig. S2. This structure is reminiscent of $NiI_2$, a well-known multiferroic material with a helimagnetic structure [30].

The X-ray refinement suggested significantly Cr-site deficiency together with homogeneous Se/Br mixing at all crystallographic Wyckoff sites. Energy dispersive X-ray (EDX) spectroscopy mapping (Supplementary Fig. S1) confirms uniform distributions of constitute elements throughout crystal. The structure is refined accordingly, with composition could be written as $Cr_{0.854}(Se_{2/7}Br_{5/7})_2$, i.e., $Cr_3Se_2Br_5$. Many synthetic attempts have been carried out to tune the Cr deficiency levels and Se / Br ratios and turn out to be all unsuccessful. All these attempts result in a precise $Cr_3Se_2Br_5$ stoichiometry, indicating this phase in fact is a line compound rather than a chemically doped system with relatively broad range of Se:Br ratios. This phenomenon further suggests this defect structure is stabilized by charge compensation associated with Cr vacancies, which balance the anion stoichiometry of the Se/Br framework.

Given the odd number of valence electrons per primitive cell, $Cr_3Se_2Br_5$ corresponds to a half-filled electronic configuration and would be expected to exhibit metallic behaviors within a simple band picture. Instead, $Cr_3Se_2Br_5$ is strongly insulating, pointing to the importance of electron correlations and indicating a correlation-driven insulating ground state, like Mott insulator. Optical absorption measurements in Fig. 1f,g further confirm the insulating character, revealing an optical gap of 1.35 eV.

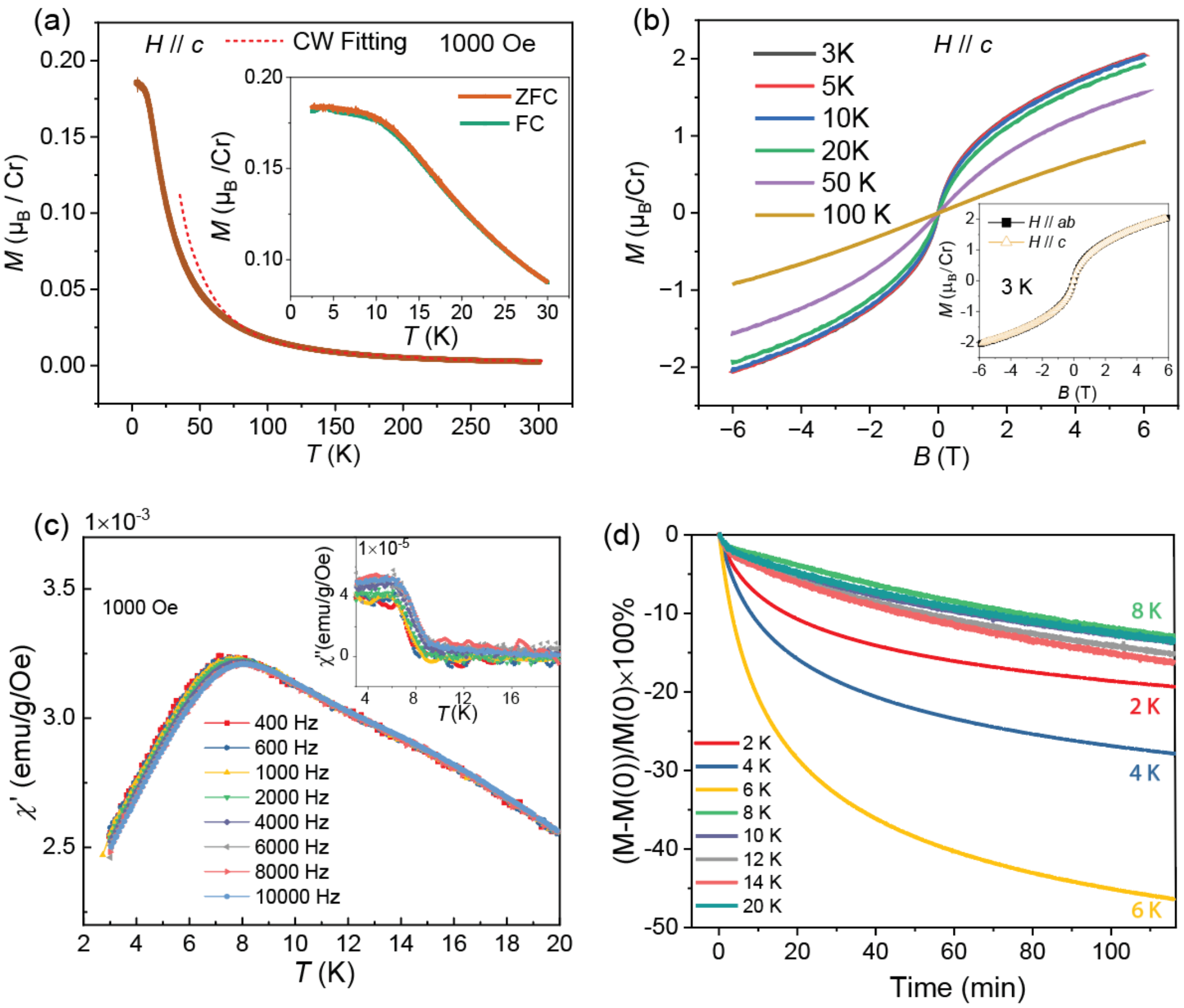

Fig. 2. **Magnetic properties of $Cr_3Se_2Br_5$** (a) Temperature dependence of the magnetic susceptibility with the external magnetic field applied along the crystallographic *c* axis. Inset: enlarged view of zero-field-cooling (ZFC) and field-cooling (FC) modes from 1.8 to 30 K. (b) Isothermal magnetization at temperatures from 3 to 100 K. Inset: M – H curve at 3 K with the magnetic field applied along the *ab* plane and the crystallographic *c* axis, highlighting the isotropic magnetic behavior. (c) Temperature dependence of the real part of the AC susceptibility under a DC magnetic field of 1000 Oe at various frequencies from 400 – 10000 Hz. Inset: imaginary part of the AC susceptibility. (d) Isothermal relaxation curves measured over the temperature range of 2–20 K.

The temperature dependence of the magnetization in both the zero-field cooled (ZFC) and field-cooled (FC) modes is presented in Fig. 2a. Magnetic susceptibility shows the absence of long-range magnetic order down to 1.8 K, which is the lowest temperature accessible in our measurements. The magnetization is fitted via the modified Curie-Weiss law, $\frac{1}{\chi-\chi_0} = \left(\frac{C}{T-\theta_{CW}}\right)^{-1}$, where $\chi_0$ is the temperature-independent contribution. The yielded $\theta_{CW} = 29\,\mathrm{K}$, indicates ferromagnetic interactions in these highly disordered triangular planes. The effective magnetic moment of $Cr^{3+}$ derived from the Curie constant C in the Curie-Weiss law is 3.9 $\mu_B$. This aligns closely with the theoretical spin-only value of 3.87 $\mu_B$ for an $S = 3/2$ state. The exchange interaction constant $J$ is estimated to be 0.22 meV (~2.5 K). Such a value is comparable to that in other triangular-lattice QSL candidates like $YbMgGaO_4$ and $Na_2BaCo(PO_4)_2$ [31–34]. In the inset of Fig. 2a, an expansion of the magnetization between 1.8 – 30 K is shown. Upon cooling, ZFC and FC curves overlapping with each, both shows a plateau like behavior below 10 K. Fig. 2b shows magnetization as a function of applied field at different temperatures. At 100 K, a straight M-H curve is observed, demonstrating the paramagnetic behavior. Upon cooling, an unusual curve without hysteresis is observed down to 1.8 K, further excluding the possibility of long-range magnetic order in $Cr_3Se_2Br_5$. It is important to note that the M-H measurements do not detect the specific transitions where plateau-like characteristics emerge. The inset of Fig. 2b exhibits the nearly identical magnetization for H // ab and H // c, highlighting the isotropic magnetic nature of layered $Cr_3Se_2Br_5$.

To further understand the origin of the plateau like M(T) curve at low temperatures, we carried out AC susceptibility measurements under various magnetic fields, as shown in Fig. 2c. The real part of the AC susceptibility exhibits a maximum at ~ 7 K, indicative of the development of short-range magnetic correlations. The observed peak shows a very weak frequency-dependence over two orders of frequency range (400 Hz to 10kHz) with a negligible shift toward higher temperatures, ruling out conventional spin-glass behavior. The imaginary component $\chi''$ , as shown in the inset of Fig. 2c, does not exhibit a well-defined peak across the measured frequency range, excluding the superparamagnetic behaviors, where a pronounced and frequency-dependent dissipation peak is expected whereas the responses here remain weak and featureless.

To further probe the magnetic dynamics, we performed TRM measurements following a field-quench protocol, as shown in Fig. 2d. For each data set, the sample has been cooled from 100 K to the specified final temperature under a 1 T field, and thermalized at such temperature for 30 minutes, the field is then turned off, and the magnetization decay is recorded as a function of time.

For an ordered system with a static moment, magnetization endures indefinitely. And for a canonical glassy system, slower spin dynamics are expected upon cooling. Our system exhibits non-monotonic and strongly temperature-dependent relaxation behavior. In particular, the relaxation is most pronounced at intermediate temperatures (~ 6 K), rather than increasing continuously toward lower temperatures. As a whole, magnetization of $Cr_3Se_2Br_5$ indicates the lack of a well-defined long-range order nor typical relaxation timescale, but instead points to a broad distribution of dynamical processes. Such behavior is consistent with a strongly disordered magnetic state with continuously distributed low-energy excitations, like valence-bond glass, a random singlet state, or a cluster glass[35] , rather than conventional glassy freezing.

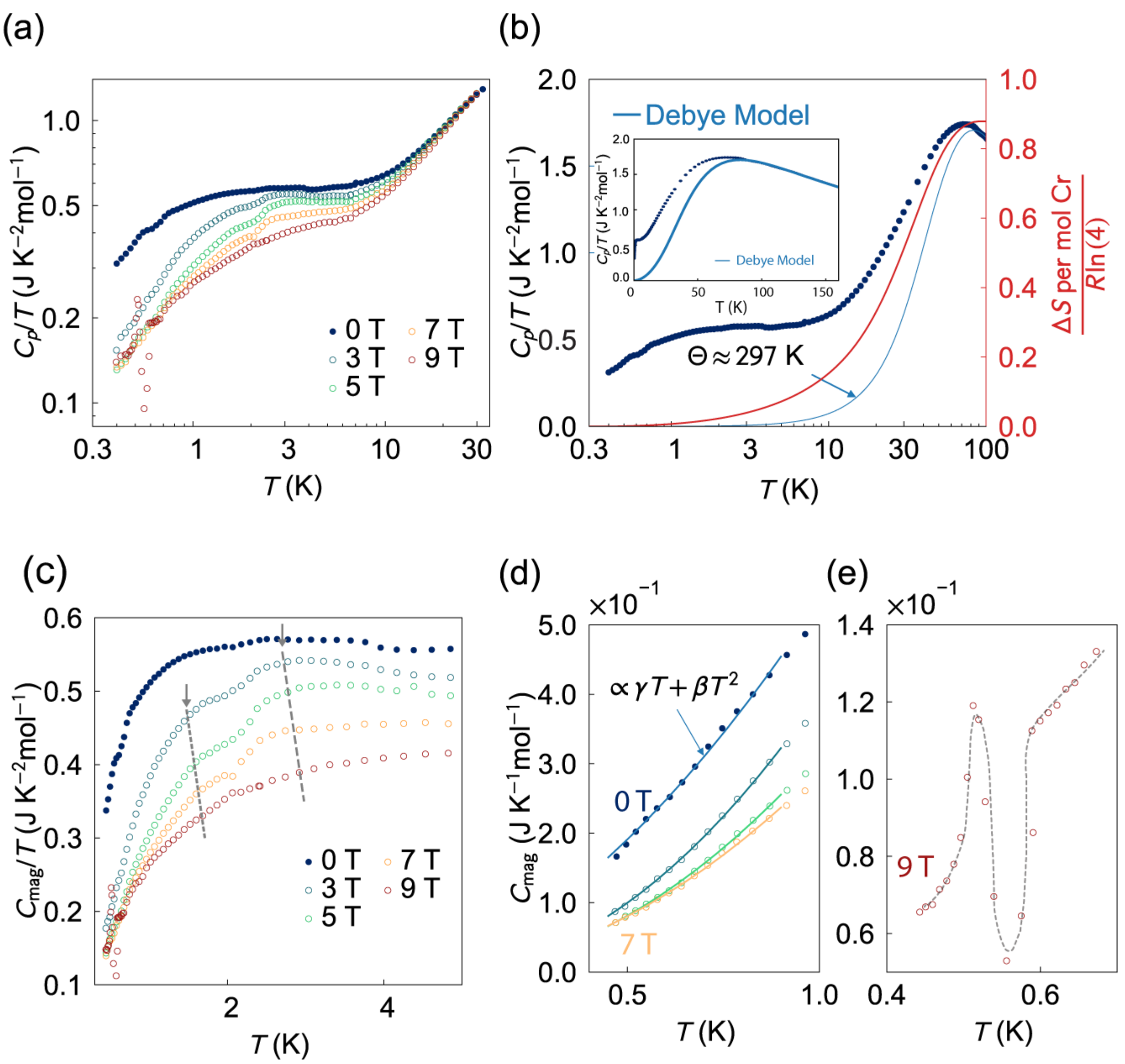

Fig. 3 **Specific heat of $Cr_3Se_2Br_5$.** (a) Low-temperature specific heat result of $Cr_3Se_2Br_5$ under zero and various magnetic fields applied along the *c* direction. (b) The fitting of zero-field heat capacity using the Debye model (blue line). Red line: The integrated entropy release from electronic spins of the $Cr^{3+}$ between 0.4 – 100 K. Inset highlights the good agreement with the Debye model at temperatures above 80 K. (c) Magnetic heat capacity of $Cr_3Se_2Br_5$ at zero and various magnetic fields. (d) Zoom-in view of the magnetic heat capacity from 0.38 – 1 K, following the $C \sim T^2$ scaling behavior. (e) Zoom-in view of the magnetic heat capacity under 9 T, showing a field-induced anomaly.

We next performed specific heat measurements to understand the low-temperature thermodynamic behavior. Fig. 3a shows the specific heat of $Cr_3Se_2Br_5$ down to 0.4 K under magnetic fields up to 9 T. At zero field, the specific heat decreases monotonically with no sharp anomaly indicating lack of long-range magnetic order or structure transition. At intermediate temperatures above 2 K, the specific heat exhibits a noticeable linear-in-$T$ contribution, corresponding to a coefficient $\gamma_n$ of 600 $mJ \cdot mol^{-1} \cdot K^{-2}$, as exhibited in Supplementary Fig. S4. The observation of such a large coefficient in a 3d-electron system is quite intriguing, and is reminiscent of heavy-fermion behavior. However, heavy-fermion behaviors are typically observed in the 4f- and 5f-electron materials, which are driven by the hybridization between itinerant conduction electrons and localized magnetic moments [36–38]. A rare exception is $LiV_2O_4$, which shows a similarly large $\gamma_n$ of 420 mJ $mol^{-1}$ $K^{-2}$ while the physical origin remains under debate[39,40].

Fig. 3b highlights the lattice contribution to the specific heat, fitted using the Debye model, where the Debye temperature is estimated to be 297 K. For an ideal $S = 3/2$ system, the magnetic entropy is estimated to be $R$ In4. After the lattice contribution subtracted, integration the magnetic entropy reaches about 85% of $R$In4. The missing entropy suggests that a significant portion of the spin degrees of freedom remain fluctuating, indicating strong quantum fluctuations beyond conventional thermal fluctuations.

In general, low temperature specific heat contains multiple contributions, including lattice, electrons, nuclear, and magnetism. Given the highly insulating nature of $Cr_3Se_2Br_5$, the electronic contribution is negligible. Furthermore, the nuclear contribution, $C_{nul} \propto AT^{-2}$ diverges and is typically dominant in systems contain heavy nuclei, such as rare-earth or 4f elements[38,41,42]. Thus, our analysis focuses exclusively on the lattice and magnetic term, $C_{tot} = C_{lat} + C_{mag}$. The magnetic specific heat, obtained after subtracting the lattice contribution, is displayed in Fig. 3c. At zero field, there is no sharp anomaly observed, indicating the absence of long-range magnetic order. Instead, some broad non-smooth feature mimicking two broad humps emerges with the centers located near 1 K and 2.5 K. The two humps are enhanced under external magnetic fields up to 9 T. A dual-peak structure in $C(T)$ has been predicted by several theoretical models for both frustrated triangular[43] and kagome Heisenbergmagnet[44]. This feature is generally attributed to the development of short-range correlations at distinct energy scales and has been discussed as a possible signature of quantum spin liquid behavior in certain regimes.

To further understand the ultra-low temperature behavior, the magnified data below 1 K are shown in Fig. 3d. Magnetic specific heat from 0 to 7 T approximately follows a $T^2$ power law. As $T$continues decrease below 1 K, $C_{mag}/T$ exhibits an additional linear-in-$T$ contribution with a finite intercept. Following recent phenomenological analyses of low-temperature specific heat in frustrated systems[45], the low-temperature magnetic specific heat in the experimentally accessible regime can be described by

$$C_{mag} = \gamma T + \beta T^2.$$

where the $T^2$ contribution captures the low-energy background, while the additional linear term improves the description at the lowest accessible temperatures

The extracted values of $\gamma$ and $\beta$ under various external fields are summarized in Supplementary Table 3. We note that the present temperature range does not yet allow an unambiguous determination of whether the finite $\gamma$ represents a true asymptotic $T \to 0$ coefficient or a crossover contribution associated with disorder-induced low-energy excitations. Nevertheless, within the experimentally accessible temperature window, the extracted $\gamma$ and $\beta$ remain reasonably robust against moderate variations of the fitting boundaries.

The $\beta$ term, remains around 300 mJ/mol/K$^3$ across measured field range with weak field dependence, indicating a robust $T^2$ scaling behavior. Such a $T^2$ behavior can, in principle, arise from phonons in a quasi-two-dimensional lattice. However, the observed field response and overall behavior suggest that the excitations are unlikely to be purely phononic. Instead, the $T^2$ scaling is consistent with gapless excitations with a linear density of states, as expected for Dirac-like dispersions in certain quantum spin liquid models[46,47].

In contrast, the linear coefficient $\gamma$ exhibits a pronounced field dependence. At zero field, $\gamma$ is roughly estimated to be 229 mJ/mol/K$^2$, which is rapidly suppressed to 10 mJ/mol/K$^2$ by a modest magnetic field of 3T, followed by slightly recovering at higher fields. The pronounced field dependence suggests that the linear-in-$T$ contribution is closely tied to magnetic degrees of freedom, potentially associated with local magnetic fluctuations. A crossover scale, following the conventional phenomenological specific contributions $T^* = \gamma/\beta \approx 0.76\ K$ is established, below which the linear contribution becomes comparable or slightly dominates to the $T^2$ term in the experimentally accessible lowest-temperature regime.

As previously noted in several triangular-lattice systems, the application of a sufficiently strong magnetic field can enhance spin correlations and, in some cases, induce field-driven magnetic ordering. While such behavior has been discussed in the context of U(1) Dirac quantum spin liquid scenarios[48], it is not unique to them and can arise more generally in frustrated magnetic systems. In $Cr_3Se_2Br_5$, the presence of strong intrinsic disorder further complicates this interpretation. Fig. 3e shows a zoomed-in view of the magnetic specific heat under 9 T from 0.4 - 1 K with the phonon contribution subtracted. Notably, the feature observed here does not resemble a traditional λ-type peak or a Schottky-like anomaly. Instead, it shows a complex morphology with a combined peak-and-dip structure. Such an unusual thermodynamic signature suggests the presence of unconventional low-energy excitations. However, its microscopic origin remains unclear at present.

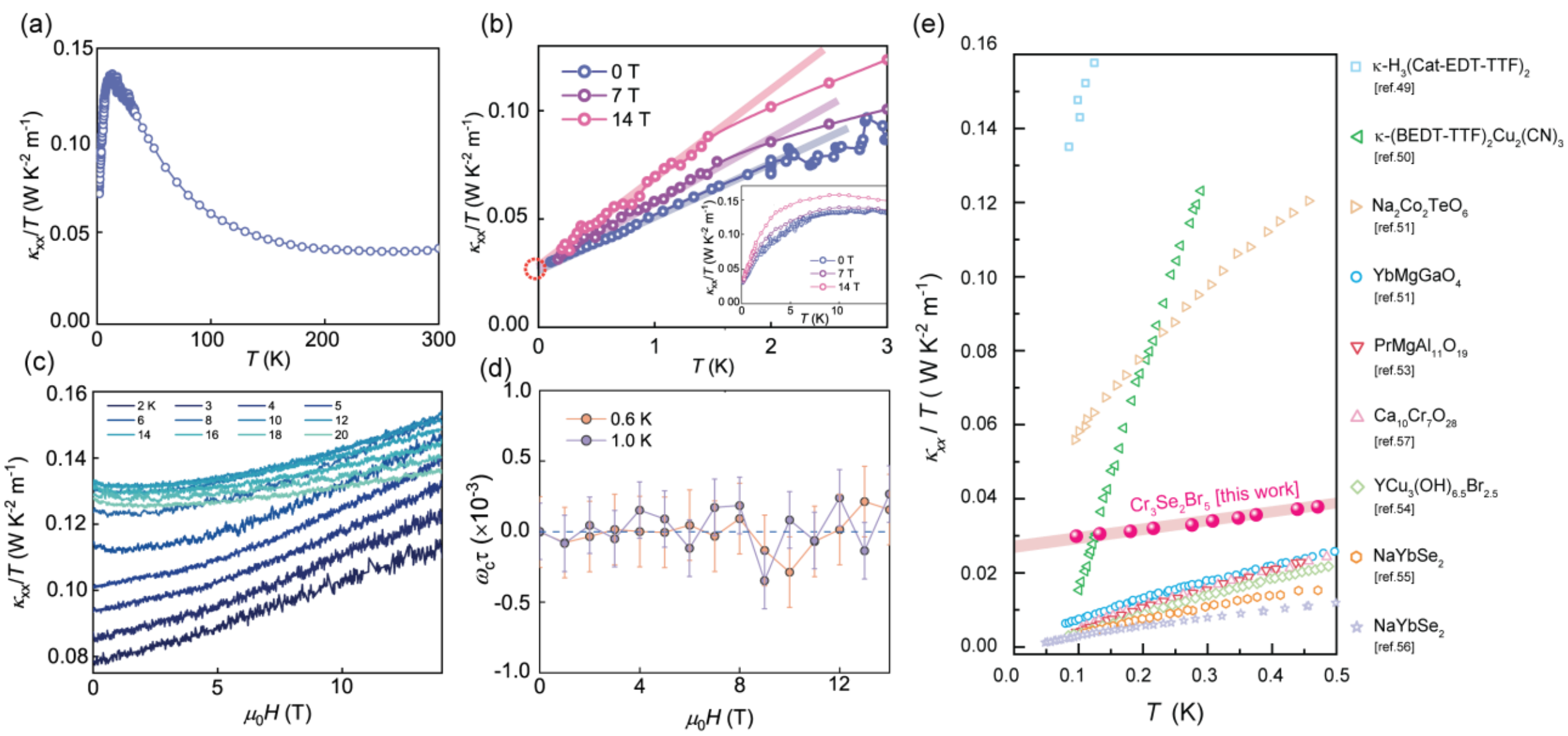

Fig. 4. **Thermal conductivity of $Cr_3Se_2Br_5$**. The heat current was applied within the crystallographic plane, perpendicular to the $c$ axis, while the external magnetic field was applied along the $c$ axis. (a) Temperature dependence of the longitudinal thermal conductivity plotted as $\kappa_{xx}/T$, showing a pronounced phonon peak near 13.5 K, indicating dominant phonon heat transport away from the lowest-temperature limit. (b) Low-temperature $\kappa_{xx}/T$ down to 0.1 K. A clear finite residual linear term of ~ 0.03 W m$^{-1}$ K$^{-2}$ is obtained from the $T \to 0$ extrapolation, consistent with gapless mobile charge-neutral excitations. (c) Magnetic-field dependence of $\kappa_{xx}/T$ above 2 K, showing an overall enhancement with field, consistent with reduced phonon scattering from magnetic excitations. (d) Thermal Hall angle, $\tan\theta_H^{\mathrm{th}} = \kappa_{xy}/\kappa_{xx}$, as a function of field at 0.6 and 1.0 K. Within the experimental resolution (~ $2 \times 10^{-4}$), no detectable thermal Hall response is observed. (e) Comparison of residual thermal conductivity of $Cr_3Se_2Br_5$ (solid symbols) with representative quantum spin liquid candidates (open symbols) near the zero-temperature limit.[49–57]

Thermal conductivity is a particularly powerful probe of mobile heat-carrying excitations, as localized excitations may contribute substantially to the specific heat but are ineffective at transporting heat over long distances[49,58,59]. Fig. 4a shows the temperature dependence of the longitudinal thermal conductivity divided by temperature $\kappa_{xx}/T$, of $Cr_3Se_2Br_5$ from 0.1 to 300 K. Upon cooling from high temperature, $\kappa_{xx}/T$ first increases and then develops a pronounced phonon peak near 13.5 K, due to the suppression of phonon – phonon Umklapp scattering followed by the eventual limitation of the phonon mean free path by sample boundaries and disorder. Away from the lowest temperature limit, heat transport in $Cr_3Se_2Br_5$ is dominated primarily by phonon[60,61].

To extract information on the low-energy excitations, we focus on the low-temperature limit of the thermal conductivity. In the presence of gapless mobile excitations, $\kappa_{xx}/T$ can retain a finite residual term as $T \to 0$. Fig. 4b displays the low-temperature behavior of $\kappa_{xx}/T$ down to 0.1 K. A clear non-zero residual is observed. Linear extrapolation of $\kappa_{xx}/T$ to $T = 0$ yields a sizeable intercept of approximately 0.03 W m$^{-1}$ K$^{2}$, which persists over the entire magnetic-field range investigated. In addition to this finite intercept, $\kappa_{xx}/T$ exhibits an approximately linear temperature dependence at low temperatures, consistent with the layered, quasi-two-dimensional character of $Cr_3Se_2Br_5$.

We further examined the field dependence of the thermal conductivity. Fig. 4c shows the magnetic field dependence of $\kappa_{xx}$ above 2 K. Overall, $\kappa_{xx}$ increases substantially with increasing field, in contrast to the field dependence of the specific heat, which decreases with field. This opposite trend suggests that magnetic excitations primarily act as scattering centers for heat-carrying phonons. As the magnetic field suppresses spin fluctuations and tends to polarize the spins, the phonon mean free path increases, leading to an enhancement of the longitudinal thermal conductivity.

Fig. 4d presents the thermal Hall response of $Cr_3Se_2Br_5$. The thermal Hall angle, defined as $\tan\theta_H^{\mathrm{th}} = \kappa_{xy}/\kappa_{xx}$, is plotted as a function of magnetic field at selected temperatures of 0.6 and 1.0 K. Within experimental uncertainty, the signal remains indistinguishable from zero, with a resolution limit of approximately $2 \times 10^{-4}$. This indicates the absence of a detectable field-induced thermal Hall effect in this temperature range. The combination of a finite $\kappa_0/T$ and a negligible thermal Hall response suggests that the itinerant excitations are mobile and gapless, yet do not exhibit a sizable transverse response associated with strong chirality or Berry-curvature effects.

Fig. 4e summarizes the thermal conductivity $\kappa_{xx}/T$ of $Cr_3Se_2Br_5$ with various quantum spin liquid candidates in the sub-Kelvin regime. Notably, the magnitude of $\kappa_{xx}/T$ in $Cr_3Se_2Br_5$ is markedly higher than that observed in majority quantum spin liquid candidates, including the triangular-lattice system $YbMgGaO_4$ and the kagome-like $YCu_3(OH)_{6.5}Br_{2.5}$.[51,54] While it remains below the exceptionally high conductivity reported for the organic insulator $\kappa$-$H_3$(Cat-EDT-TTF)$_2$ and the honeycomb compound $Na_2Co_2TeO_6$,[49,52] the robust and well-defined linear extrapolation in $Cr_3Se_2Br_5$ provides strong evidence for mobile low-energy excitations. This intermediate magnitude, together with the finite residual term, places $Cr_3Se_2Br_5$ in a regime where heat is carried by mobile yet disorder-influenced excitations.

Lastly, we discuss the nature of these low-energy excitations. Currently, there is no satisfactory explanation for all our observations. The strong intrinsic disorder in $Cr_3Se_2Br_5$ causing diversity in the local interactions, makes it challenging to identify the microscopic origin of the low-energy excitations quantitatively. Moreover, the strong suppression of the phenomenological $\gamma$ term under modest magnetic fields seems to contrast with the nearly field-independent residual $\kappa_0/T$ up to 14 T, suggesting that the dominant thermodynamic low-energy spectral weight and the heat-carrying channel are not identical. While the specific heat reflects the total low-energy density of states, including both localized and non-localized excitations, thermal conductivity selectively probes only the mobile component. The strong field dependence of $\gamma$ therefore indicates that a substantial fraction of the low-energy spectral weight is associated with localized or quasi-localized magnetic excitations associated with disorder, magnetic domains, or low-energy magnetic fluctuations, whose spectral weight can be rapidly suppressed through Zeeman splitting or partial spin polarization. In contrast, the residual thermal conductivity appears to arise from a small subset of more spatially extended neutral excitations that remain mobile and robust against magnetic field.

In the following, we focus on the origin of this exotic itinerant component. The absence of long-range magnetic order suggests that conventional magnon excitations are not well-defined. More importantly, even if short-range magnetic excitations persist, they cannot account for the observed finite residual thermal conductivity $\kappa_0/T$ as T → 0, which requires itinerant, gapless excitations. Given that $Cr_3Se_2Br_5$ is highly insulating, contribution from mobile charge carriers can be excluded, implying that heat is transported by charge-neutral excitations.

The presence of a finite $\kappa_0/T$ further distinguishes this system from conventional glassy behavior. In amorphous and glassy systems, thermal conductivity at low temperatures typically follows universal scaling laws and $\kappa/T \rightarrow 0$ as T → 0. In contrast, $Cr_3Se_2Br_5$ exhibits a sizable residual linear term that lies well outside the glassy regime[62–64], as shown in Supplementary Fig. S5, indicating a qualitatively different transport mechanism. In frustrated insulating magnets, such behavior is often associated with fractionalized quasiparticles, such as spinons in QSL states. However, the strong intrinsic disorder and the relatively large spin value in $Cr_3Se_2Br_5$ make a clean QSL scenario less evident. Alternative scenarios, such as spin-ice states, are also unlikely given the triangular lattice geometry and the absence of strong Ising anisotropy.

The intrinsic disorder naturally raises the possibility of disorder-induced low-energy excitations. Rather than acting solely as a perturbation that suppresses quantum coherence, disorder has been theoretically proposed to reshape the low-energy excitation spectrum and stabilize partially itinerant neutral excitations. For example, in random-singlet or valence-bond-glass scenarios, spatial randomness may generate connected singlet networks through mechanisms such as "entanglement retiling", enabling heat transport without charge[15,56]. Such a scenario is qualitatively consistent with the observations of coexistence of strong intrinsic disorder, absence of long-range magnetic order, and finite residual thermal conductivity in $Cr_3Se_2Br_5$, suggesting a disorder-modified quantum-disordered state with partially mobile neutral excitations.

Alternative scenarios involve the interplay between disorder and lattice degrees of freedom. While phonons are typically the dominant heat carriers in insulators, additional low-energy excitations can emerge in strongly disordered systems. In conventional structural glasses, tunneling two-level systems (TLS) act primarily as scatterers and cannot account for a finite $\kappa_0/T$. However, TLS of electronic or magnetic origin in a disordered Mott insulator, especially when strongly hybridized with phonons, may behave differently[65,66]. In such a scenario, hybridization between localized modes and extended lattice vibrations could give rise to additional diffusive or even itinerant modes that contribute directly to both specific heat and thermal transport. At present, the origin of the itinerant heat carriers in $Cr_3Se_2Br_5$ remains an open question, and may involve unconventional magnetic excitations, disorder-induced modes, or an interplay between the two.

In summary, we have observed striking phenomena where mobile charge-neutral low-energy excitations in a strongly disordered triangular-lattice Mott candidate $Cr_3Se_2Br_5$. Structural refinement reveals a network of edge-sharing $Cr^{3+}$ octahedra with approximately 15% Cr-site deficiency and homogeneous Se/Br site mixing within a geometrically frustrated triangular lattice. Combined magnetic, thermodynamic, and thermal transport measurements indicate the presence of mobile charge-neutral excitations in this highly insulating system at low temperatures. More broadly, our results place $Cr_3Se_2Br_5$ in an unusual regime where strong intrinsic disorder, frustration, and electronic correlations collectively reshape low-energy excitations. These findings suggest that neutral itinerancy can persist even in heavily disordered high-spin correlated insulators, extending beyond the conventional clean $S = 1/2$ QSL paradigm.

## ACKNOWLEDGMENTS

Work at the University of California, Berkeley and Lawrence Berkeley National Laboratory was funded by the U.S. DOE, Office of Science, Office of Basic Energy Sciences, Materials Sciences and Engineering Division under Contract No. DE-AC02-05CH11231 (Quantum Materials Program KC2202). Work at UT Dallas was funded by US Air Force Office of Scientific Research (AFOSR) (FA9550-19-1-0037), National Science Foundation(NSF) (2324033 and 2516364) and Office of Naval Research (ONR) (N00014-23-1-2020). Work at the University of Michigan is supported by the National Science Foundation under Award No. DMR-2317618 (transport measurements), by the U.S. Department of Energy under Award No. DE-SC0020184 (magnetometry). Work at the Air Force Research Laboratory, Foundational Technology Directorate, was funded by AFOSR grant 26RXCOR010.

## REFERENCES:

1. Wölfle, P. Quasiparticles in condensed matter systems. *Rep. Prog. Phys.* **81**, 032501 (2018).

2. Anderson, P. W. More Is Different. *Science* **177**, 393–396 (1972).

3. Kitaev, A. Anyons in an exactly solved model and beyond. *Ann. Phys.* **321**, 2–111 (2006).

4. Senthil, T. & Fisher, M. P. A. $Z_2$ gauge theory of electron fractionalization in strongly correlated systems. *Phys. Rev. B* **62**, 7850–7881 (2000).

5. Baskaran, G., Zou, Z. & Anderson, P. W. The resonating valence bond state and high-Tc superconductivity — A mean field theory. *Solid State Commun.* **63**, 973–976 (1987).

6. Anderson, P. W. Resonating valence bonds: A new kind of insulator? *Mater. Res. Bull.* **8**, 153–160 (1973).

7. Balents, L. Spin liquids in frustrated magnets. *Nature* **464**, 199–208 (2010).

8. Zhou, Y., Kanoda, K. & Ng, T.-K. Quantum spin liquid states. *Rev. Mod. Phys.* **89**, 025003 (2017).

9. Semeghini, G. *et al.* Probing topological spin liquids on a programmable quantum simulator. *Science* **374**, 1242–1247 (2021).

10. Lee, P. A., Nagaosa, N. & Wen, X.-G. Doping a Mott insulator: Physics of high-temperature superconductivity. *Rev. Mod. Phys.* **78**, 17–85 (2006).

11. Niggemann, N., Hering, M. & Reuther, J. Classical spiral spin liquids as a possible route to quantum spin liquids. *J. Phys.: Condens. Matter* **32**, 024001 (2019).

12. Wen, J., Yu, S.-L., Li, S., Yu, W. & Li, J.-X. Experimental identification of quantum spin liquids. *npj Quantum Mater.* **4**, 12 (2019).

13. Savary, L. & Balents, L. Quantum spin liquids: a review. *Rep. Prog. Phys.* **80**, 016502 (2016).

14. Scheie, A. O. *et al.* Proximate spin liquid and fractionalization in the triangular antiferromagnet $KYbSe_2$. *Nat. Phys.* **20**, 74–81 (2024).

15. Kimchi, I., Nahum, A. & Senthil, T. Valence Bonds in Random Quantum Magnets: Theory and Application to $YbMgGaO_4$. *Phys. Rev. X* **8**, 031028 (2018).

16. Zhu, Z., Maksimov, P. A., White, S. R. & Chernyshev, A. L. Disorder-Induced Mimicry of a Spin Liquid in YbMgGaO4. *Phys. Rev. Lett.* **119**, 157201 (2017).

17. Murayama, H. *et al.* Effect of quenched disorder on the quantum spin liquid state of the triangular-lattice antiferromagnet 1T−$TaS_2$. *Phys. Rev. Res.* **2**, 013099 (2020).

18. Yu, Y. J. *et al.* Heat transport study of the spin liquid candidate 1T-$TaS_2$. *Phys. Rev. B* **96**, 081111 (2017).

19. Liu, W. *et al.* Electrical transport crossover and large magnetoresistance in selenium deficient van der Waals $HfSe_{2-x}$ ($0\leq x\leq 0.2$). *Phys. Rev. Mater.* **8**, 054006 (2024).

20. Li, S. *et al.* Transport anomalies in the layered compound BaPt4Se6. *npj Quantum Mater.* **6**, 80 (2021).

21. Miranda, E. & Dobrosavljević, V. Disorder-driven non-Fermi liquid behaviour of correlated electrons. *Rep. Prog. Phys.* **68**, 2337 (2005).

22. Sheng, Q. & Henley, C. L. Ordering due to disorder in a triangular Heisenberg antiferromagnet with exchange anisotropy. *J. Phys.: Condens. Matter* **4**, 2937 (1992).

23. Shimokawa, T., Watanabe, K. & Kawamura, H. Static and dynamical spin correlations of the S=1/2 random-bond antiferromagnetic Heisenberg model on the triangular and kagome lattices. *Phys. Rev. B* **92**, 134407 (2015).

24. Kawamura, H. & Uematsu, K. Nature of the randomness-induced quantum spin liquids in two dimensions. *J. Phys.: Condens. Matter* **31**, 504003 (2019).

25. Furukawa, T. *et al.* Quantum Spin Liquid Emerging from Antiferromagnetic Order by Introducing Disorder. *Phys. Rev. Lett.* **115**, 077001 (2015).

26. Savary, L. & Balents, L. Disorder-Induced Quantum Spin Liquid in Spin Ice Pyrochlores. *Phys. Rev. Lett.* **118**, 087203 (2016).

27. Buessen, F. L., Hering, M., Reuther, J. & Trebst, S. Quantum Spin Liquids in Frustrated Spin-1 Diamond Antiferromagnets. *Phys. Rev. Lett.* **120**, 057201 (2018).

28. Khait, I., Stavropoulos, P. P., Kee, H.-Y. & Kim, Y. B. Characterizing spin-one Kitaev quantum spin liquids. *Phys. Rev. Res.* **3**, 013160 (2021).

29. Xu, C. *et al.* Possible Kitaev Quantum Spin Liquid State in 2D Materials with S=3/2. *Phys. Rev. Lett.* **124**, 087205 (2020).

30. Song, Q. *et al.* Evidence for a single-layer van der Waals multiferroic. *Nature* **602**, 601–605 (2022).

31. Shen, Y. *et al.* Evidence for a spinon Fermi surface in a triangular-lattice quantum-spin-liquid candidate. *Nature* **540**, 559–562 (2016).

32. Li, Y. *et al.* Gapless quantum spin liquid ground state in the two-dimensional spin-1/2 triangular antiferromagnet $YbMgGaO_4$. *Sci. Rep.* **5**, 16419 (2015).

33. Li, F.-Y., Li, Y.-D., Yu, Y., Paramekanti, A. & Chen, G. Kitaev materials beyond iridates: Order by quantum disorder and Weyl magnons in rare-earth double perovskites. *Phys. Rev. B* **95**, 085132 (2017).

34. Zhong, R., Gao, T., Ong, N. P. & Cava, R. J. Weak-field induced nonmagnetic state in a Co-based honeycomb. *Sci. Adv.* **6**, eaay6953 (2020).

35. Day, R. P. *et al.* Colossal magnetoresistance and anisotropic spin dynamics in the antiferromagnetic semiconductor $Eu_5Sn_2As_6$. *Phys. Rev. B* **111**, 054406 (2025).

36. Löhneysen, H. v., Rosch, A., Vojta, M. & Wölfle, P. Fermi-liquid instabilities at magnetic quantum phase transitions. *Rev. Mod. Phys.* **79**, 1015–1075 (2007).

37. Si, Q. & Steglich, F. Heavy Fermions and Quantum Phase Transitions. *Science* **329**, 1161–1166 (2010).

38. Stewart, G. R. Heavy-fermion systems. *Rev. Mod. Phys.* **56**, 755–787 (1984).

39. Kondo, S. *et al.* $LiV_2O_4$: A Heavy Fermion Transition Metal Oxide. *Phys. Rev. Lett.* **78**, 3729–3732 (1997).

40. Chmaissem, O., Jorgensen, J. D., Kondo, S. & Johnston, D. C. Structure and Thermal Expansion of LiV2O4: Correlation between Structure and Heavy Fermion Behavior. *Phys. Rev. Lett.* **79**, 4866–4869 (1997).

41. Fujita, T., Satoh, K., Ōnuki, Y. & Komatsubara, T. Specific heat of the dense Kondo compound CeCu6. *J. Magn. Magn. Mater.* **47**, 66–68 (1985).

42. Fulde, P. & Jensen, J. Electronic heat capacity of the rare-earth metals. *Phys. Rev. B* **27**, 4085–4094 (1983).

43. Wang, Y. R. Specific heat of quantum Heisenberg model on a triangular lattice with two exchange parameters and its application to He3 adsorbed on graphite. *Phys. Rev. B* **45**, 12608–12611 (1992).

44. Singh, R. R. P. & Oitmaa, J. High-temperature series expansion study of the Heisenberg antiferromagnet on the hyperkagome lattice: Comparison with Na4Ir3O8. *Phys. Rev. B* **85**, 104406 (2012).

45. Zheng, G. *et al.* Thermodynamic Evidence of Fermionic Behavior in the Vicinity of One-Ninth Plateau in a Kagome Antiferromagnet. *Phys. Rev. X* **15**, 021076 (2025).

46. Wen, X.-G. Quantum orders and symmetric spin liquids. *Phys. Rev. B* **65**, 165113 (2002).

47. Ran, Y., Hermele, M., Lee, P. A. & Wen, X.-G. Projected-Wave-Function Study of the Spin-1/2 Heisenberg Model on the Kagomé Lattice. *Phys. Rev. Lett.* **98**, 117205 (2007).

48. Bordelon, M. M. *et al.* Field-tunable quantum disordered ground state in the triangular-lattice antiferromagnet $NaYbO_2$. *Nat. Phys.* **15**, 1058–1064 (2019).

49. Shimozawa, M. *et al.* Quantum-disordered state of magnetic and electric dipoles in an organic Mott system. *Nat. Commun.* **8**, 1821 (2017).

50. Yamashita, M. *et al.* Thermal-transport measurements in a quantum spin-liquid state of the frustrated triangular magnet $\kappa$-$(BEDT\text{-}TTF)_2Cu_2(CN)_3$. *Nat. Phys.* **5**, 44–47 (2009).

51. Rao, X. *et al.* Survival of itinerant excitations and quantum spin state transitions in $YbMgGaO_4$ with chemical disorder. *Nat. Commun.* **12**, 4949 (2021).

52. Hong, X. *et al.* Phonon thermal transport shaped by strong spin-phonon scattering in a Kitaev material $Na_2Co_2TeO_6$. *npj Quantum Mater.* **9**, 18 (2024).

53. Tu, C. P. *et al.* Gapped quantum spin liquid in a triangular-lattice Ising-type antiferromagnet $PrMgAl_{11}O_{19}$. *Phys. Rev. Res.* **6**, 043147 (2024).

54. Hong, X. *et al.* Heat transport of the kagome Heisenberg quantum spin liquid candidate $YCu_3(OH)_{6.5}Br_{2.5}$: Localized magnetic excitations and a putative spin gap. *Phys. Rev. B* **106**, L220406 (2022).

55. Zhu, Z. *et al.* Fluctuating magnetic droplets immersed in a sea of quantum spin liquid. *Innov.* **4**, 100459 (2023).

56. Lyu, Y. *et al.* Entanglement Randomness and Gapped Itinerant Carriers in a Frustrated Quantum Magnet. *Phys. Rev. X* **15**, 041035 (2025).

57. Ni, J. M. *et al.* Ultralow-temperature heat transport in the quantum spin liquid candidate $Ca_{10}Cr_7O_{28}$ with a bilayer kagome lattice. *Phys. Rev. B* **97**, 104413 (2018).

58. Yamashita, M. *et al.* Highly Mobile Gapless Excitations in a Two-Dimensional Candidate Quantum Spin Liquid. *Science* **328**, 1246–1248 (2010).

59. Li, N. *et al.* Possible itinerant excitations and quantum spin state transitions in the effective spin-1/2 triangular-lattice antiferromagnet $Na_2BaCo(PO_4)_2$. *Nat. Commun.* **11**, 4216 (2020).

60. Acharyya, P. *et al.* Glassy thermal conductivity in $Cs_3Bi_2I_6Cl_3$ single crystal. *Nat. Commun.* **13**, 5053 (2022).

61. Agne, M. T., Hanus, R. & Snyder, G. J. Minimum thermal conductivity in the context of diffuson -mediated thermal transport. *Energy Environ. Sci.* **11**, 609–616 (2018).

62. Anderson, P. w., Halperin, B. I. & Varma, c. M. Anomalous low-temperature thermal properties of glasses and spin glasses. *Philos. Mag.: A J. Theor. Exp. Appl. Phys.* **25**, 1–9 (1972).

63. Phillips, W. A. Tunneling states in amorphous solids. *J. Low Temp. Phys.* **7**, 351–360 (1972).

64. Tavakoli, A. *et al.* Universality of thermal transport in amorphous nanowires at low temperatures. *Phys. Rev. B* **95**, 165411 (2017).

65. Schechter, M. & Stamp, P. C. E. Inversion symmetric two-level systems and the low-temperature universality in disordered solids. *Phys. Rev. B* **88**, 174202 (2013).

66. Vojta, T. Disorder in Quantum Many-Body Systems. *Annu. Rev. Condens. Matter Phys.* **10**, 1–20 (2018).

67. Liu, W. *et al.* A Three-Stage Magnetic Phase Transition Revealed in Ultrahigh-Quality van der Waals Bulk Magnet CrSBr. *ACS Nano* **16**, 15917–15926 (2022).